\documentclass[aps,prb,twocolumn,showpacs,superscriptaddress,preprintnumbers,amsmath,amssymb,longbibliography]{revtex4-1}

\usepackage[utf8]{inputenc}
\usepackage{dcolumn}
\usepackage{bm}
\usepackage{ulem}    
\usepackage{xcolor}
\usepackage{gensymb}   
\usepackage{tikz}
\usepackage[pdftex,colorlinks=true,pdfstartview=FitV,linkcolor=blue,citecolor=blue,urlcolor=blue]{hyperref}
\begin{document}
\title{Size effect on the structural and magnetic phase transformations of iron nanoparticles}

\author{Alexis Front}
\affiliation{Laboratoire d'Etude des Microstructures, ONERA-CNRS, UMR104, Universit\'e Paris-Saclay, BP 72, Ch\^atillon Cedex, 92322, France}
\affiliation{Department of Chemistry and Materials Science, Aalto
University, 02150 Espoo, Finland}
\affiliation{Department of Applied physics, Aalto
university, FI-00076 Aalto, Espoo, Finland}
\author{Georg Daniel F\"orster}
\affiliation{Interfaces, Confinement,Mat\'eriaux et Nanostructures (ICMN), CNRS, Universit\'e d'Orl\'eans, 45071, Orl\'eans, France}
\author{Chu-Chun Fu}
\affiliation{Universit\'e Paris-Saclay, CEA, Service de recherche en Corrosion et Comportement des Mat\'eriaux, SRMP, F-91191 Gif-sur-Yvette, France}
\author{Cyrille Barreteau}
\affiliation{Universit\'e Paris-Saclay, CEA, CNRS, SPEC, 91191, Gif-sur-Yvette, France}
\author{Hakim Amara}
\email{hakim.amara@onera.fr}
\affiliation{Laboratoire d'Etude des Microstructures, ONERA-CNRS, UMR104, Universit\'e Paris-Saclay, BP 72, Ch\^atillon Cedex, 92322, France}
\affiliation{Universit\'e de Paris, Laboratoire Mat\'eriaux et Ph\'enom\`enes Quantiques (MPQ), F-75013, Paris, France}

\date{\today}

\begin{abstract}
 
Iron nanoparticles are among the most promising low-dimensional materials in terms of applications. This particularity is attributable to the magnetic properties of these nanoparticles, which exhibit different allotropes as a function of temperature. In this work, we sought to characterise at the atomic scale how their structural and magnetic transformations can be affected by the size. To achieve this objective, we developed a tight-binding model incorporating a magnetic contribution via a Stoner term implemented in a Monte Carlo code to relax the structure and the magnetic state. Using our approach, we show that magnetism is strongly reinforced by the surface, which leads to increase the Curie temperature as the size of the particle decreases contrary to the solid-solid transition temperature. Our work thus provides a deep understanding at the atomic scale of the key factors that determines the structural and magnetic properties of Fe nanoparticles, shedding more light on their unique character which is crucial for further applications.

\end{abstract}


\maketitle


\section{Introduction}

Iron is a very abundant element on Earth, and its use and mastery has been a milestone in the history of humanity giving rise to the Iron Age period. This particularity is mainly due to its unique versatility, which derives largely from its polymorphic nature. Indeed, one of the main characteristics of Fe is its highly diverse and complex phase diagram, with a number of closely correlated structural and magnetic phase transitions~\cite{Hasegawa1983, Besson1990, Wang2023}. More precisely, its thermodynamic phase diagram presents four solid phases in addition to the liquid one~\cite{Strong1973, Swartzendruber1982}. Furthermore, iron transits  from  the  body-centered  cubic (BCC) $\alpha$ phase to the face-centered cubic (FCC) $\gamma$ phase around 1200 K. In this context, a detailed understanding of the mechanisms of $\alpha$ (ferrite) to $\gamma$ (austenite) transformation of iron is of primary importance in materials science for various applications, but also for the fundamental theory of phase transitions in solids~\cite{Kormann2016, Magomedov2021}. However, the mechanisms underlying this structural transformation are not so trivial since it is closely related to a transition from a ferromagnetic (FM) to a paramagnetic (PM) state. 

In addition, Fe nanoparticles (NPs) are an important class of nanomaterials offering high potential as magnetic recording media, catalyst for steam reforming or even green material to prevent environmental pollution~\cite{Huber2005, Saif2016, Bolade2020} due to their high surface area to volume ratio leading to unique physical features. More specifically, these effects are directly due to a heterogeneous surface presenting different surface sites with specific properties which are preponderant for small sizes~\cite{Perini1984, Martin1991, Henry2005}. Compared to bulk counterparts, size reduction can have a major impact on physical properties such as melting temperature~\cite{Buffat1976}, surface energies~\cite{Molleman2018, Amara2022} as well as mechanical properties~\cite{Feruz2016, Erbi2023}. This can be particularly important in the case of iron NPs if structural transformations and magnetic properties differ from the bulk impacting their use in a range of applications. A typical example is the growth of carbon nanotubes from iron nanoparticles, where the final structure of the tubes, and therefore their electronic properties, depends on the physico-chemical state of the Fe catalyst and above all its crystallographic structure~\cite{Ding2004, Harutyunyan2008, He2018, Magnin2018, Hu2023}. However, we still lack a reliable quantitative understanding of phase transformations in case of Fe NPs. More precisely, it is well known that the magnetism plays a key role in different characteristics of bulk iron in its various crystal structures~\cite{Herper1999, Mrovec2011}. Regarding Fe NPs, the impact of magnetism on phase transformations is not obvious and deserves more studies on this topic. 

To address this issue, large-scale atomic simulations are perfectly suited although they are generally still limited by the transferability of interatomic potentials~\cite{Meiser2020} and, in the particular case of iron, have to explicitly take into account the magnetic degree of freedom. In this respect, efforts to model iron-based materials at the atomic scale have been carried out, both to improve fundamental understanding of different properties and to guide the development of materials for specific applications~\cite{Chiesa2011, Mrovec2011}. Many studies on bulk iron have therefore revealed that thermodynamic properties such as $\alpha$ to $\gamma$ phase transformation can be captured using interatomic potentials including magnetism, which plays a crucial role~\cite{Okatov2009, Lavrentiev2010, Ma2017, Wang2023}. To the best of our knowledge, no study has been devoted to characterising the phase transformation of iron nanoparticles at the atomic scale. From an experimental point of view, High-Resolution Transmission Electron Microscopy (HRTEM) analysis have shown that Fe NPs adopt the BCC structure at room temperature. Unfortunately no phase transition has been studied contrary to other chemical elements from \textit{in situ} TEM observations or other techniques~\cite{Barnard2009, Liu2021}. Meanwhile, previous theoretical works based on phenomenological interatomic potentials have shown that the NP's size has a significant effect on phase transformations such as $\alpha$ to $\gamma$ structures~\cite{Li2003, Sandoval2009} or melting properties~\cite{Shibuta2007, Shibuta2008, Fedorov2017}. In all cases, atomic relaxations are of course considered whereas the magnetic contribution is implicitly considered since the fitting of the model parameters is based on spin-polarised DFT calculations. In addition, there are numerous models on the variation of magnetism with NP size, which are mainly based on effective models from rigid lattice~\cite{Souto2016, Santos2020} or DFT calculations~\cite{Eone2019}. Furthermore, the magnetic moments of metallic NP and especially Fe at different temperatures have been studied with contradictory conclusions regarding the dependence of the Curie temperature (T$_{\mathrm{C}}$) on the size of the NP. According to Heisenberg model by application of a mean field approximation, it is commonly admitted that T$_{\mathrm{C}}$ is related to the average coordination number~\cite{Jensen2006}. As a result, it is expected to decrease in case of thin films compared to the bulk counterpart as well as with decreasing NP size. However, radically opposite conclusions are achieved when studies are based on electronic structure calculations. Within this framework, the magnetic order is reinforced by the surface~\cite{Polak1995} since the reduction in coordination number leads to a sharper density of state~\cite{Ducastelle1991}. Consequently, the T$_{\mathrm{C}}$ is expected to increase as the size of the NPs decreases. Such a dependence has been reported experimentally for Co, Ni and Fe NPs of very small sizes~\cite{Billas1993, Billas1994}.

In this work, we investigate how the structural (solid to solid) and magnetic (FM to PM) transitions are correlated and their dependence on the size of Fe NPs. Our study is carried out with a magnetic interatomic potential based on the tight-binding (TB) framework that we specifically developed for this purpose. It is based on the fourth-moment approximation for the band term enabling some of the magnetic characteristics of iron to be captured successfully with the simplest possible model in terms of moments. This potential is coupled to a Monte Carlo (MC) code in the canonical ensemble where trials of atomic displacements as well as fluctuations of local magnetic moments are performed to reach the thermodynamic equilibrium state of the studied systems. We show that this approach is perfectly suited to deal with large Fe NP systems (up to $\sim$ 5 nm) and thus revealing the influence of their size on the solid to solid transition. Furthermore, the detailed analysis of the local magnetic moment helps to understand how its strengthening at the surface in case of small NPs has a major influence on the evolution of T$_{\mathrm{C}}$.

\section{Results and discussion}

\subsection{Bulk iron system}

\begin{figure}[htbp!]
  \includegraphics[width=1\linewidth]{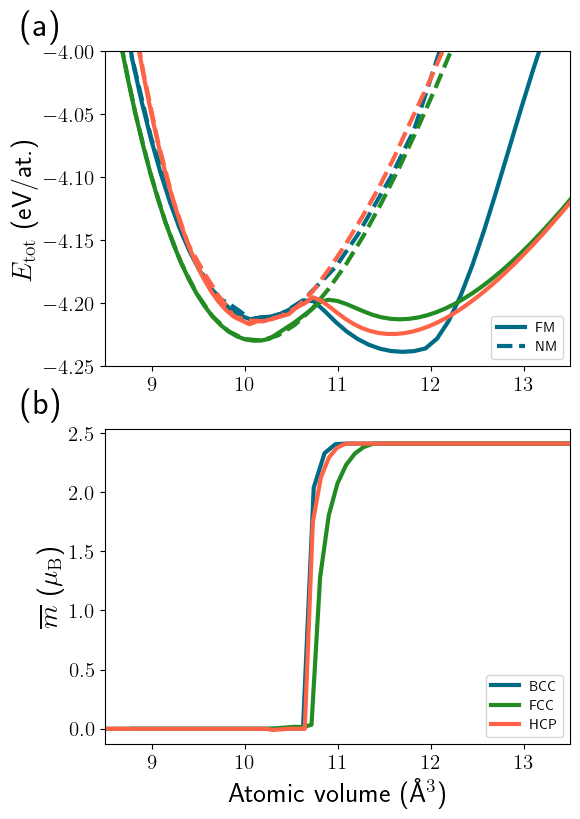}
  \caption{(a) Total energy ferromagnetic (FM) and nonmagnetic (NM) calculations and (b) total magnetic moment as a function of atomic volume for BCC (blue), FCC (green), and HCP (orange) at 0 K.}
    \label{fig:Figure_1}
\end{figure}

\begin{figure}[htbp!]
  \includegraphics[width=1\linewidth]{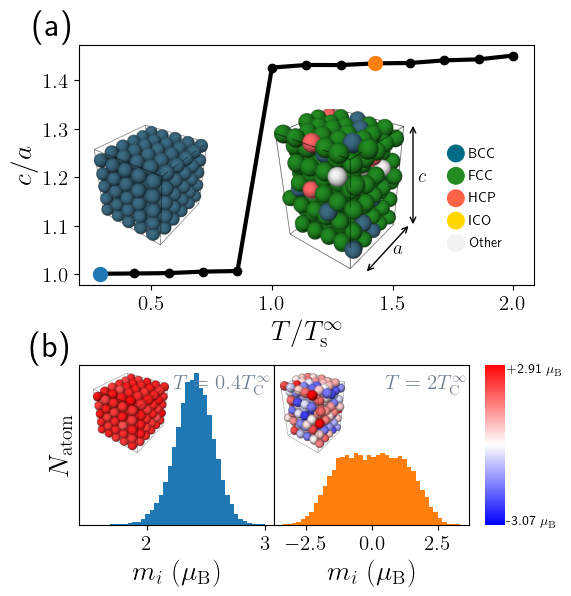}
  \caption{(a) Structural evolution as a function of temperature of a bulk Fe system containing 250 atoms. (b) Distribution of local magnetic moments in the ferromagnetic (at 0.4T$_{\mathrm{C}}^{\infty}$) and paramagnetic (at 2.0T$_{\mathrm{C}}^{\infty}$) regimes.}
    \label{fig:Figure_2}
\end{figure}

Figures~\ref{fig:Figure_1} and ~\ref{fig:Figure_2} provide two examples illustrating the ability of our TB model in characterising the properties of Fe at 0 K but also at finite temperature. 

First, the relative stability of the various phases, as well as the influence of magnetism on the system can be determined from the energy versus atomic volume curves plotted in Figure~\ref{fig:Figure_1}a. As already discussed~\cite{Autes2006, Paxton2008, Mrovec2011, Paxton2014}, ferromagnetism clearly plays a major role in determining the stability of bulk Fe structures. Our TB model confirms that the FM BCC phase is found to be the ground state while the FCC phase is the most stable one in the case of non-magnetic calculations. Moreover, our TB model predicts a magnetic moment equal to 2.41 $\mu_{\mathrm{B}}$ for the BCC structure which is close to the 2.31 $\mu_{\mathrm{B}}$ values given by DFT calculations or 2.22 $\mu_{\mathrm{B}}$ from experiments~\cite{Fu2004}. To go further in the validation and transferability of the interatomic potential for studying thermodynamic properties, it is fundamental to assess its reliability at finite temperatures. A particular challenge in our study of Fe NPs is to derive an interatomic potential able to describe solid to solid phase transformation and the Curie temperature. Before considering the complex case of a NP, it is therefore judicious to assess our TB model on a bulk system. In the following a BCC system containing 250 Fe atoms is considered from which off-lattice MC simulations are performed including atomic relaxations, lattice expansion as well as magnetic moment fluctuations.

To characterize the phase transformation, we then performed heating MC simulations i.e. the simulation at the next temperature starts from the previous converged configuration of the last temperature. Moreover, to characterise the structural transition, box relaxations are also included corresponding to MC trials on the amplitude of the box along its three directions while constraining an orthogonal structure. From these calculations, it is then possible to estimate the Curie temperature T$_{\mathrm{C}}^{\infty}$ where the magnetization drops from 2.41 $\mu_{\mathrm{B}}$ to zero (see Fig. S2 of the Supplemental Materials). Furthermore, the structural evolution is monitored through the variation of the $c/a$ value with the temperature. The results are presented in Figure~\ref{fig:Figure_2}. At $T<T_{\mathrm{C}}^{\infty}$ the system is of the BCC FM type ($c/a=1$) to reach a FCC PM crystal at 
$T>T_{\mathrm{C}}^{\infty}$ ($c/a=\sqrt{2}$). To gain an insight at the atomic level, the structure identification is performed based on the Ackland-Jones analysis~\cite{Ackland2006} using the \texttt{Ovito} software~\cite{ovito}. As shown in Figure~\ref{fig:Figure_2}, there is a phase transformation from a BCC crystal to a system containing mainly atoms in a FCC environment. To extend this analysis, the distribution of local magnetic moments in the ferromagnetic and paramagnetic regimes are displayed in Figure~\ref{fig:Figure_2}b. At low temperature, a Gaussian-type distribution is observed around the magnetic moment value of the fundamental state (2.41 $\mu_{\mathrm{B}}$). In this ferromagnetic state, the dispersion is simply due to thermal fluctuations. After the complete phase transformation, the distribution of the magnetic moment is broader and its amplitude is weaker corresponding to the paramagnetic state. An exhaustive list of tests for the validation of the potential highlighting the transferability and robustness of our potential are presented in Sec. III of the Supplemental Material.

\subsection{Phase transformations of Fe NP}

\begin{figure}[htbp!]
  \includegraphics[width=0.90\linewidth]{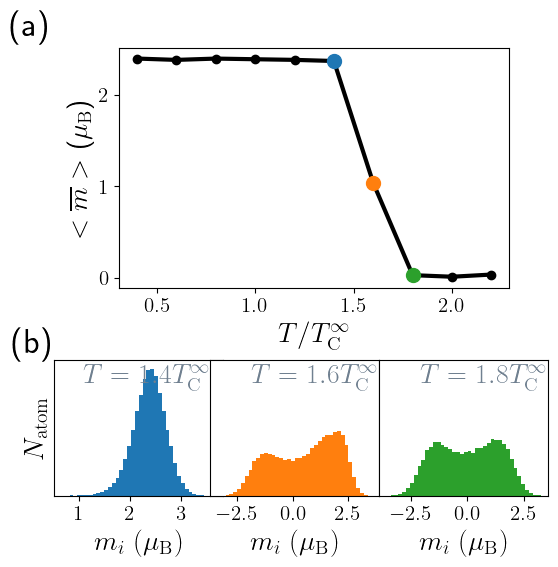}
  \caption{Magnetic properties of a Fe NP containing 671 atoms. (a) Total magnetic moment average as a function of temperature. (b) Local magnetic moment distribution in histogram forms at different temperatures.}
    \label{fig:Figure_3}
\end{figure}
\begin{figure}[htbp!]
  \includegraphics[width=1\linewidth]{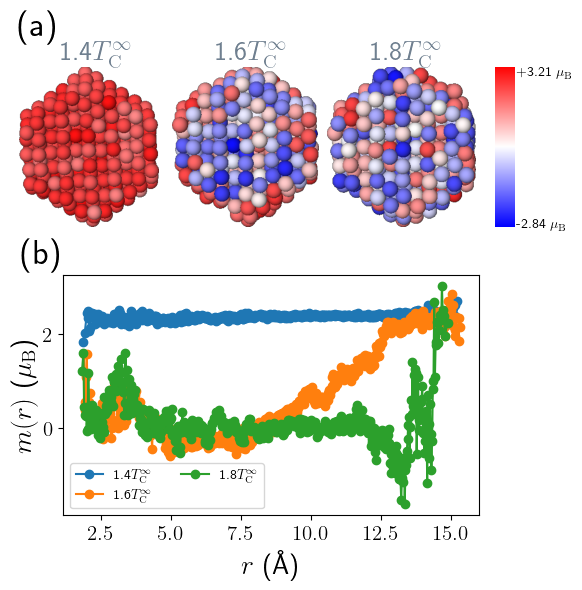}
  \caption{(a) Analysis of the local magnetic moment within the Fe NP containing 671 atoms where slice views are presented for different temperatures. (b) Local magnetic moment as a function of the distance from the center of the Fe NP containing 671 atoms at different temperatures.}
    \label{fig:Figure_4}
\end{figure}
The next step is to investigate at the atomic level the magnetic and structural phase transitions of Fe NPs. In the present work, the structures considered are based on the BCC lattice as observed experimentally at room temperature~\cite{Huber2005,Carpenter2001, Carvell2010}. To highlight size effects, NPs with diameters ranging from 20 to 50 \AA~(175 to 2459 atoms, respectively) are taken into account. To begin, we focus on a BCC NP containing 671 atoms and analyse its structural and magnetic arrangement evolution as a function of the temperature by performing simulated annealing.  First, the Curie temperature is calculated and presented in Figure~\ref{fig:Figure_3}a, estimated to be around 1.8T$_{\mathrm{C}}^{\infty}$ and therefore higher than the value obtained for a bulk system. Furthermore, the local magnetic distribution, presented in a histogram form, shows that the system goes smoothly from a rather sharp monomodal state to a wider distribution when crossing from FM to PM state (see Figure~\ref{fig:Figure_3}b). As already known~\cite{Front2022}, such feature is observed for transition metals with high $d$ band filling, in contrast to lower fillings, where a bimodal distribution is obtained in the PM state. Interestingly, we can note that during the transition (T= 1.6T$_{\mathrm{C}}^{\infty}$), the bimodal distribution is not symmetrical, which tends to suggest that two populations are present.

To go deeper in this analysis, we consider the evolution of the local magnetic moments distribution within the NP during the increase in temperature. As seen in Figure~\ref{fig:Figure_4}a, the initial configuration exhibits atoms with identical magnetic moments saturated at the value of 2.41 $\mu_{\mathrm{B}}$. Unfortunately, our model, based on a simple description of the electronic structure, is not capable of capturing the exaltation of the local magnetic moment due to the surface, already discussed in the literature via DFT calculations at 0 K~\cite{Eone2019, Souto2016}. A fifth or sixth moment approximation of the electronic density of states would be better but rather expensive in term of CPU calculations~\cite{Glanville1988, Nastar1995}. Nevertheless, this discrepancy does not prevent us from qualitatively decrypting the magnetic properties of Fe NPs since it reveals the influence of surfaces on the reinforcement of magnetism in case of finite temperature calculations. 
\begin{figure}[htbp!]
  \includegraphics[width=1\linewidth]{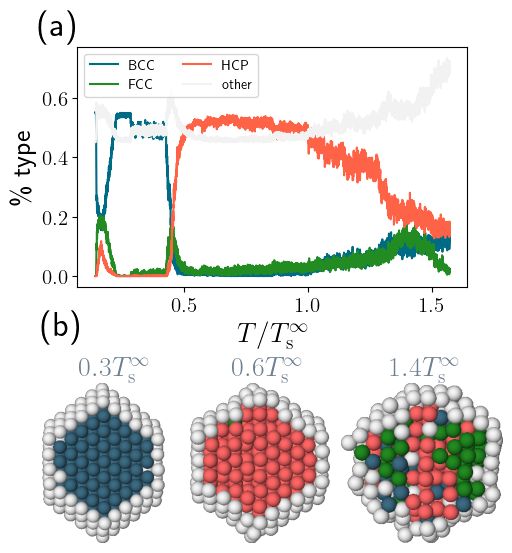}
  \caption{Structural evolution of a Fe NP containing 671 atoms based on the Ackland-Jones analysis. (a) Average configurations as a function of the temperature. (b) Analysis of the local structure where slice views are presented for different temperatures.}
    \label{fig:Figure_5}
\end{figure}
Indeed, when we look at the local evolution of the magnetic moments presented in Figure~\ref{fig:Figure_4}a for different temperatures, a number of very pertinent conclusions can be drawn. 
\begin{figure*}[htbp!]
\includegraphics[width=1.0\linewidth]{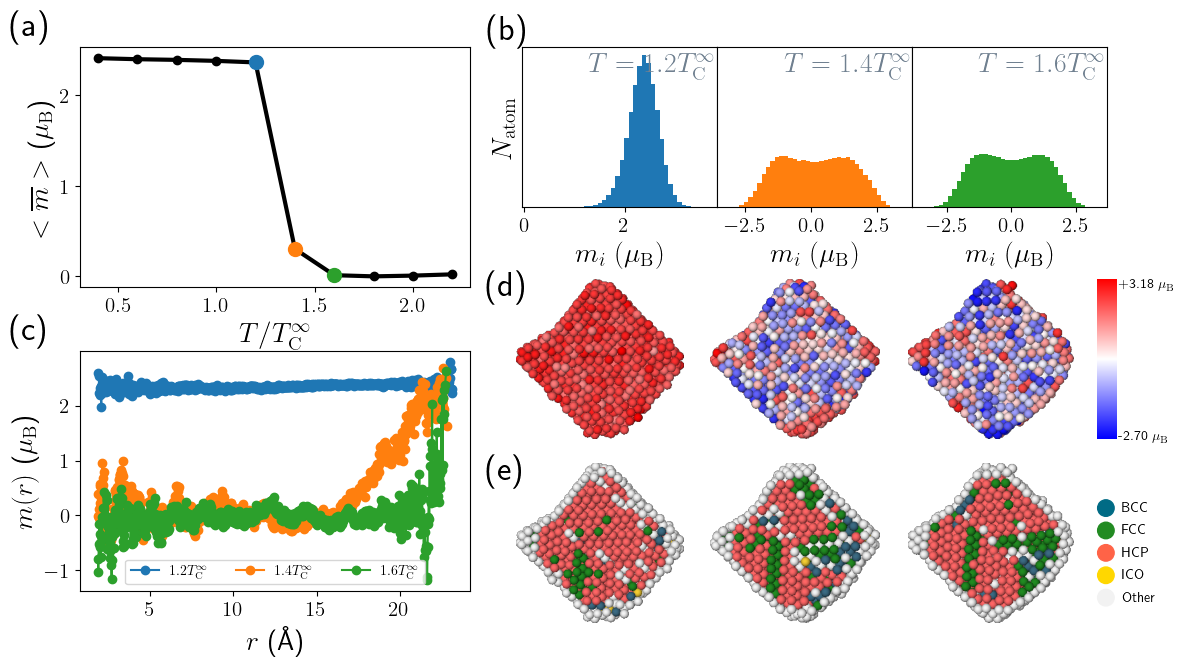}
\caption{Magnetic and structural evolution of a Fe NP containing 2459 atoms. (a) Total magnetic moment average as a function of the temperature. (b) Local magnetic moment distribution in histogram forms at different temperatures. (c) Local magnetic moment as a function of the distance from the center of the NP at different temperatures. (d) Analysis of the local magnetic moment within the NP Fe where slice views are presented for different temperature. (e) Analysis of the local structure where slice views are presented for different temperatures. }
\label{fig:Figure_6}
\end{figure*}
As the temperature rises, fluctuations in the magnetic moment inside the NP are observed. Above the T$_{\mathrm{C}}$ ($T>1.6 T_{\mathrm{C}}^{\infty}$), the range of amplitudes of the local magnetic moment is very large with a total sum equal to zero, corresponding to a paramagnetic state. Interestingly, a visual inspection reveals that the first atoms to be affected by the magnetic change as the temperature increases are the bulk ones. Surface atoms, on the other hand, remain stable up to a threshold temperature of around 1.6T$_{\mathrm{C}}^{\infty}$. The analysis of the local magnetism profiles along the radius of the NP presented in Figure~\ref{fig:Figure_4}b confirms this remark in a more quantitative way and provides a further step forward. Note that this analysis is based on an average of the local magnetic moments in a layer of a given thickness and for a set of MC configurations. At low temperatures, the local magnetic moment is constant around a value of $\sim$ 2.30 $\mu_{\mathrm{B}}$. When the temperature rises (T = 1.6 T$_{\mathrm{C}}^{\infty}$), surface atoms have strong $m_{i}$, while for some core atoms (located up to 7.5 \AA~, i.e. 3 layers) this quantity drops to zero. Then the surface atoms are also affected by the increase in temperature until they reach an average value close to zero. This trend is quite expected since it is known that the local spin magnetic moment can be magnified due to the decrease in coordination as shown by DFT calculations at 0 K in case of infinite surfaces~\cite{Soulairol2010} as well as NPs~\cite{Eone2019,Souto2016}. An important consequence of this is that the presence of surface atoms in Fe NPs tends to delay the magnetic transition and thus increase the T$_{\mathrm{C}}$ compared to a bulk system. This is in agreement with previous surface studies highlighting a Curie temperature enhancement due to strengthened magnetic order at the surface~\cite{Polak1995}.

\subsection{Structural phase transition of Fe NP}

Figure~\ref{fig:Figure_5} presents the equilibrium configurations obtained from MC simulations of the NP in form of slice views to display the structural type of each particle. Initially, two populations are present: a surface characterised by vertexes, edges and (110) facets, and a core composed of the BCC type. When increasing the temperature, we can clearly identify the occurrence of HCP atoms within the structure itself at the expense of BCC-type atoms. At higher temperatures, a complete phase transition is obtained with a majority of FCC and HCP atoms and only $\sim$10 \% of BCC atoms. The presence of both HCP and FCC atoms is not so surprising insofar as the two structures are almost degenerate (less than 0.05 eV/at). Moreover, previous work based on empirical calculations without taking magnetism into account has already shown the presence of these two crystalline phases in the case of nanoparticles~\cite{Li2003,Lummen2004} or nanowires~\cite{Sandoval2009}, as well as experimentally for systems with quasi-degenerate HCP and FCC structures~\cite{Ricolleau1999}. 

\subsection{Size effect on the structural and magnetic phase transformations of Fe NPs}

Consequently, our in-depth study of an iron NP reveals that the FM to PM transition propagates from the core of the NP towards the surface. Insofar as the surface area to volume ratio varies with the size of the NP, we then sought to better establish the dependence of the structural and magnetic transitions on the size. In this context, we consider a BCC NP bigger than in the previous case which contains 2459 atoms with a diameter around 5 nm. First, we notice in Figure~\ref{fig:Figure_6}a that the Curie temperature decreases with size, as expected due to the smaller contribution of surface magnetism. As seen in Figure~\ref{fig:Figure_6}c, the analysis of the local magnetism profiles along the radius of the NP clearly shows that the contribution of the bulk is much more important than that of the surface. More precisely, at 1.4T$_{\mathrm{C}}^{\infty}$ the magnetic reinforcement is strongly localised at the surface ($\sim$ 3 \AA~thickness) with few atoms involved compared to the number of atoms considered as having a bulk behaviour within the NP ($\sim$ 20 \AA~thickness).
\begin{figure}[htbp!]
  \includegraphics[width=1\linewidth]{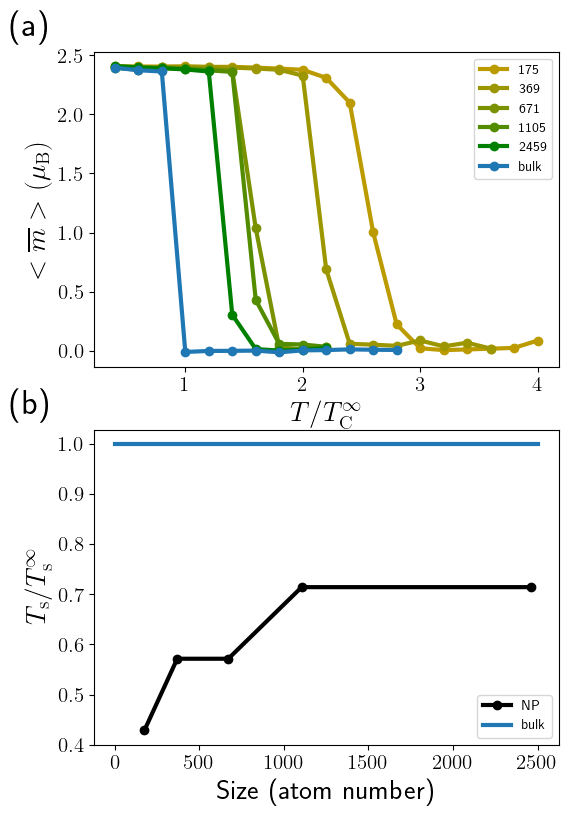}
  \caption{(a) Magnetization curves for Fe NP of different sizes. (b) Solid-solid transition temperature of iron nanoparticles at different temperatures.}
    \label{fig:Figure_7}
\end{figure}
In contrast, the structural transition temperature is increased by around 20\% compared with the nanoparticle containing 671 atoms (as seen in Figure~\ref{fig:Figure_7}b). More generally, we find a mechanism widely discussed and supported in the literature, establishing that the high surface-to-volume ratio specific to NPs strongly controls their structural properties~\cite{Ferrando2008}. A typical example is the melting temperature of NP where the formation of a liquid surface skin is a precursor effect of the solid-liquid transition~\cite{Buffat1976,Ercolessi1991, Front2023}. Therefore the surface/volume ratio variation determines the decrease of melting temperature with decreasing cluster size.  
\begin{figure}[htbp!]
  \includegraphics[width=1\linewidth]{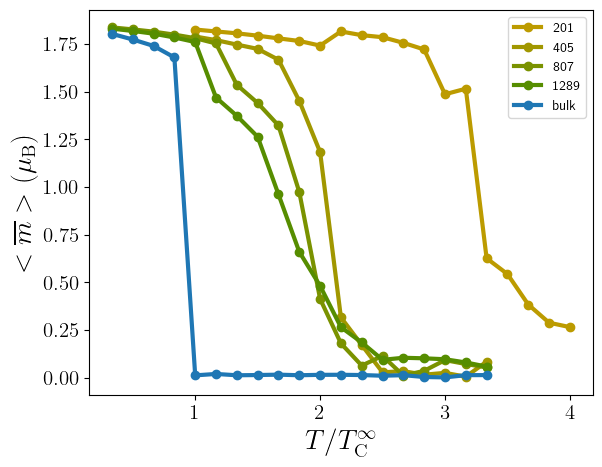}
  \caption{Magnetization curves for Co FCC NP of different sizes.}
    \label{fig:Figure_8}
\end{figure}
Finally, we subsequently extended and generalised this study to a range of NPs with diameters varying from 20 to 50 \AA. Results are presented in Figure~\ref{fig:Figure_7} and point out that the solid-solid transition temperature increases with size, contrary to T$_{\mathrm{C}}$. Although theoretical studies agree with this trend~\cite{Nowak2005,Santos2020}, many of them suggest an opposite tendency where the Curie temperature increases with the NP size~\cite{Rausch2009,Penny2019,Santos2020}. However, it is important to point out that no study is based on an atomic-scale model that explicitly includes magnetism and allows atomic relaxation. Also, at the experimental level, numerous studies based on SQUID measurements indicate that the Curie temperature increases with the size~\cite{Carpenter2001, Carvell2010}. Nevertheless, a complete and exhaustive work based on molecular beam deflection measurements clearly shows that for small NPs (close to the sizes considered here) the Curie temperature decreases when the size of the NP increases~\cite{Billas1993, Billas1994}. This conclusion is supported by an analysis of several elements (Fe, Co and Ni) while mentioning that the structural transition peculiar to iron makes the analysis of the results delicate. Similarly to this study and in order to generalise our conclusions, we also considered Co NPs which have the advantage of not being subjected to structural transformations, unlike iron. This therefore enables us to probe directly the dependence of the magnetic transition on the size of the nanoparticles. More details of the TB model for Co are given in~\cite{Front2022}. As can be seen in the Figure~\ref{fig:Figure_8}, the Curie temperature decreases with the size of the NP, which strongly suggests the robustness and generalisation of our results.

\section{Methodology}

\subsection{Magnetic tight-binding model}

The interaction between iron atoms is treated within a semi-empirical TB model~\cite{Ducastelle1991,Pettifor1995} where only $d$ orbitals are taken into account~\cite{Ducastelle1970}. We employ the recursion method to calculate the local density of electronic states $n_{i}(E)$ at all atomic sites $i$~\cite{Haydock1972,Gaspard1973}. In addition, the magnetic contribution is introduced via the Stoner model~\cite{Stoner1939} by considering the presence of local exchange fields in the band energy term giving rise to two spin populations within the collinear approximation. The technical and theoretical aspects concerning our TB model to handle the magnetic cobalt system and its extension to other transition metals as well as its transferability are provided in~\cite{Amara2009, Front2022}. A brief description is also given in Sec. I of the Supplemental Material. 

Meanwhile, it has been necessary to extend this cobalt-specific magnetic TB model to the case of iron, involving the adjustment of parameters to reproduce several bulk physical properties. All the difficulty is to define the relevant quantities specific to Fe, both from a structural and magnetic point of view, for the development of an interatomic potential with a high transferability to study phase transformation of magnetic Fe NPs. In the present work, the TB parameters have been fitted on experimental data and density functional theory (DFT) calculations using the Vienna \textit{ab initio} Simulation Package (VASP) code~\cite{Kresse1999} to reproduce the lattice parameters, the cohesive energies, the elastic moduli (bulk modulus and the shear moduli) and the magnetism state of $\alpha$ and $\beta$ bulk Fe phases at 0 K. More details about the description of the Fe-Fe interaction can be found in Sec. II of the Supplemental Material along with the resulting parameter values in Table S1. This magnetic TB model relies on local (atomic) energy calculations based on the recursion method. It is coupled with Monte Carlo (MC) simulations in order to relax the structures at given temperature where each trial corresponds to randomly choosing an atom and its displacement as well as its local magnetic moment. 

\section{Conclusions}

To conclude, we have investigated the structural and magnetic evolution of Fe NPs by means of a specifically developed TB interatomic potential including a magnetic contribution. This unique approach has enabled us to characterise in details the structural and magnetic properties of iron nanoparticles as a function of their size. Using our model based on collinear magnetism, we clearly show that the presence of the surface strongly contributes to magnetic reinforcement and therefore to an increase in the Curie temperature, in line with experimental results. Furthermore, another challenge is to study the influence of non-collinear spin calculations~\cite{Wang2023} on the properties studied here since magnetic evolution and structural effects can be strongly coupled. This is currently in progress. Finally, the present work contributes to a better understanding of the rich physics that exists at the nanoscale, paving the way to a rational search for better design of Fe NPs. \\

\section*{Supplemental material : \\Size effect on the structural and magnetic phase transformations of iron nanoparticles}

\textbf{Sec. I. Magnetic tight-binding model}\\

In the following, a brief description of the key features of our interatomic potential for treating magnetic transition metals and its transferability is provided. For the technical and theoretical aspects, a detailed account of the TB model is given in the following references~\cite{Amara2009, Front2022} while the fitting of the parameters to reproduce the energetic properties of Fe is presented. \\

In our study, the interaction between iron atoms is treated within the semi-empirical tight-binding model~\cite{Ducastelle1991,Pettifor1995} where only $d$ bands are taken into account. As in non-magnetic systems, the total energy of an atom $i$ is split in two parts, a band structure term that describes the formation of an energy band when atoms are assembled and a repulsive term which empirically reflects the ionic and electronic repulsions. We employ the recursion method to calculate the local density of electronic states $n_{i}(E)$ at all sites~\cite{Amara2009,Los2011}. Exact calculations are made of only the first four continued fraction coefficients, ($a_{1}, b_{1}, a_{2}, b_{2}$) corresponding to the first four moments of the local density of states. In addition, the magnetic contribution is introduced via the Stoner model~\cite{Stoner1939,Pettifor1995} by considering the physical presence of local exchange fields in the band energy term giving rise to two spin populations within the collinear approximation. Besides, the fourth moment approximation (FMA) is a good compromise for describing the structural properties of transition metals~\cite{Amara2009,Front2022} while having a minimal description of the density of states necessary to take into account local magnetic on-site levels and thus define two spin populations. Interestingly, the FMA model is highly effective to enable a linear scaling of CPU working time as a function of system size. This magnetic TB model relies on local (atomic) energy calculations is coupled with Monte Carlo (MC) simulations in order to relax the structures where each trial corresponds to randomly choosing an atom and its displacement as well as its local magnetic moment. By performing this procedure several times, it becomes possible to determine the equilibrium properties of iron nanoparticles of various sizes in terms of both position and magnetic state. \\

\textbf{Sec. II. Iron tight-binding model}\\

In case of transition metals, the electronic structure is defined by a narrow $d$ band hybridizing with a wider $sp$-band corresponding to nearly free electrons. Given that the cohesion properties of Fe are mainly driven by $d-d$ bonding, it is only necessary to include $d$ orbitals in the spin-polarised TB framework~\cite{Ducastelle1970}. In our $d$ band model, the Slater-Koster parameters characterizing the hopping integrals ($dd\sigma$, $dd\pi$ and $dd\delta$) are chosen according to the ratio -2:1:0 and to decrease exponentially with the following distance dependence $r$ between atoms: 
\begin{equation}
dd\lambda(r)=dd\lambda_{0}\exp\left[-q\left(\frac{r}{r_{0}}-1\right)\right] \quad,
\label{hopping_integrals}
\end{equation}
where $\lambda = \sigma, \pi, \delta$ and $q$ a parameter to be fitted. Regarding the repulsive term, a Born-Mayer expression has been adopted involving two additional parameters ($A$ and $p$):
\begin{equation}
E^{i}_{\mathrm{rep}}=A\sum_{j\neq i}\exp\left[-p\left(\frac{r_{ij}}{r_{0}}-1\right)\right]
\label{repulsif}
\end{equation}
To capture all magnetic effects, the Stoner exchange integral $I$ is the only additional parameter to be included in our TB model. This results in an energy contribution of $ -\frac{I}{4}m_{i}^{2}$~\cite{Ford2014} where $m_{i}$ is the spin moment in $\mu_{B}$ units. Indeed, $m_{i} = N_{i}\uparrow-N_{i}\downarrow$ with $N_{i}\uparrow$ and $N_{i}\downarrow$, respectively the number of electrons in majority and minority spin bands of an atom $i$. To get an efficient interatomic potential, the parameters ($dd\sigma$, $q$, $A$, $p$ and $I$) and the number of electrons $N_{d}$ have to be adjusted to reproduce several bulk physical properties of Fe. 
\begin{table}[htbp!]
\begin{center}
\setlength{\tabcolsep}{2pt}
\begin{tabular}{c|c|c|c|c|c|c}
\hline
\hline
$dd\sigma$&$q$&$r_{0}$&$A$&$p$&$I$&$N_{d}$ \\
\hline
1.08 & 3.29 & 2.42 & 0.166 & 10.5 & 1.15 & 7.59  \\
\hline
\hline
\end{tabular}
\end{center}
\caption{Fe parameters for the magnetic TB-FMA model, obtained by fitting to DFT reference data. $dd{\sigma}$, $A$ and $I$ are in eV. $r_{0}$ is in~\AA. }
\label{tab:parameters}
\end{table}

\begin{table*}[htbp!]
\begin{center}
\setlength{\tabcolsep}{2pt}
\begin{tabular}{c|c|c|c||c|c}
\multicolumn{2}{c|}{}   & NM & NM &FM &FM \\
\multicolumn{2}{c|}{}   & DFT & TB-FMA &DFT & TB-FMA \\
\hline    BCC & $a$ (\AA)  & 2.76 & 2.73 &  2.83 & 2.86 \\
    & $E_{\mathrm{coh}}$ (eV/at.)  & -3.81 & -4.21 &  -4.28 & -4.24 \\
    & $C_{11}$, $C_{12}$, $C_{44}$ (GPa)  & 89, 351, 186 & 219, 272, 87 &  278, 144, 97 & 157, 143, 95 \\
\hline   FCC & $a$ (\AA)  & 3.45 & 3.43 & 3.48 & 3.58 \\
   & $E_{\mathrm{coh}}$ (eV/at.)  & -4.12 & -4.23 & -4.13 & -4.21 \\
    & $C_{11}$, $C_{12}$, $C_{44}$ (GPa)  & 430, 223, 244 & 306, 233, 81 & 318, 127, 178 & 179, 132, 58 \\
  \hline
\end{tabular}
\end{center}
\caption{DFT and TB calculations of physical properties for non magnetic and magnetic bcc and fcc systems at 0 K.}
\label{tab:0K_E_I}
\end{table*}
All the difficulty is to define the relevant quantities specific to Fe, both from a structural and magnetic point of view, for the development of an interatomic potential with a high degree of transferability to study phase transformation of magnetic Fe NPs. In the present work, the TB parameters have been fitted on experimental data and density functional theory (DFT) calculations using the Vienna \textit{ab initio} Simulation Package (VASP) code~\cite{Kresse1999} to reproduce the lattice parameter, the cohesive energy, the elastic moduli (bulk modulus and the two shear moduli) and the magnetism state of $\alpha$ and $\beta$ bulk Fe phases at 0 K. The resulting parameter values are given in Table~\ref{tab:parameters}.\\

\textbf{Sec. III. Validity and transferability of the tight-binding model}\\

The TB parameters have been fitted on experimental values for the FCC and BCC structures namely the lattice parameter, the cohesive energy and the elastic moduli (bulk modulus and the two shear moduli) as well as their magnetic properties. All results are presented in Table~\ref{tab:0K_E_I}.

\begin{figure}[htbp!]
  \includegraphics[width=0.80\linewidth]{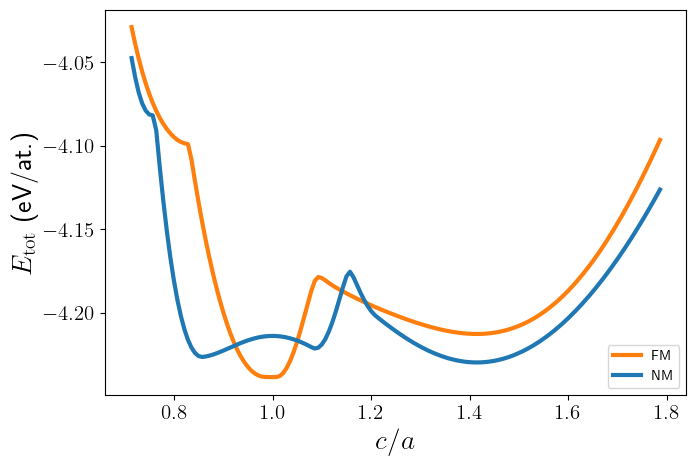}
  \caption{Bain path between the BCC and FCC phases in case of NM and FM calculations.}
    \label{fig:Figure_S1}
\end{figure}

As already discussed~\cite{Herper1999,Mrovec2011}, ferromagnetism clearly plays a major role in determining the stability of bulk Fe structures. It is interesting to note that our model successfully reproduces the main trends of this specific physics. Indeed, when magnetism is not taken into account, the $\beta$ structure is the most stable. Such behavior of non-magnetic calculations has been highlighted in previous calculations~\cite{Autes2006,Mrovec2011}. Nevertheless, magnetic calculations correct this and reproduce the experimentally stable phase, i.e. the ground-state FM $\alpha$-Fe structure which is also successfully predicted by the TB potential. Moreover, regarding the dependencies of magnetic moments as function of the lattice parameters, our TB results are in agreement with the DFT calculations in particular the increase in magnetic moment as the structures are expanded. To highlight how the ground state is driven by the magnetism, the analysis of elastic constants ($C_{ij}$) is very relevant. In case of the cubic phase, three independent elastic constants ($C_{11}$, $C_{12}$ and $C_{44}$) have to be considered, or even their combination giving rise to the tetragonal shear modulus, $C'=(C_{11}-C_{12})/2$, and the bulk modulus, $B=(C_{11}+2C_{12})/3$. Note that a negative value means that the system is mechanically unstable. As seen in Table~\ref{tab:0K_E_I}, they are calculated for non-magnetic and ferromagnetic BCC and FCC iron from our TB model, and compared with experimental as well as DFT results. It is immediately striking that the general trends are perfectly reproduced by the TB model. More specifically, it predicts correctly negative values of $C'$ for the NM BCC structure and a positive one for the ferromagnetic BCC confirming its stability respect to the tetragonal distortion. This particular behaviour is perfectly illustrated in the analysis of the energy along the Bain transformation path connecting the BCC ($c/a=1$) and FCC ($c/a=\sqrt2$) structures and presented in Figure~\ref{fig:Figure_S1}. 
Although the energy difference between the ferromagnetic and non-magnetic configurations is underestimated by the TB model, the overall trend of the energy profile is in good agreement with previous DFT calculations~\cite{Mrovec2011}. Consequently, it can be seen that the TB model is not only well suited to discriminate between the different magnetic phases of Fe, but also has the ability to predict quite subtle characteristics, such as the link between the magnetism and the structural stability of the BCC phase under tetragonal deformation. 
\begin{figure}[htbp!]
  \includegraphics[width=0.80\linewidth]{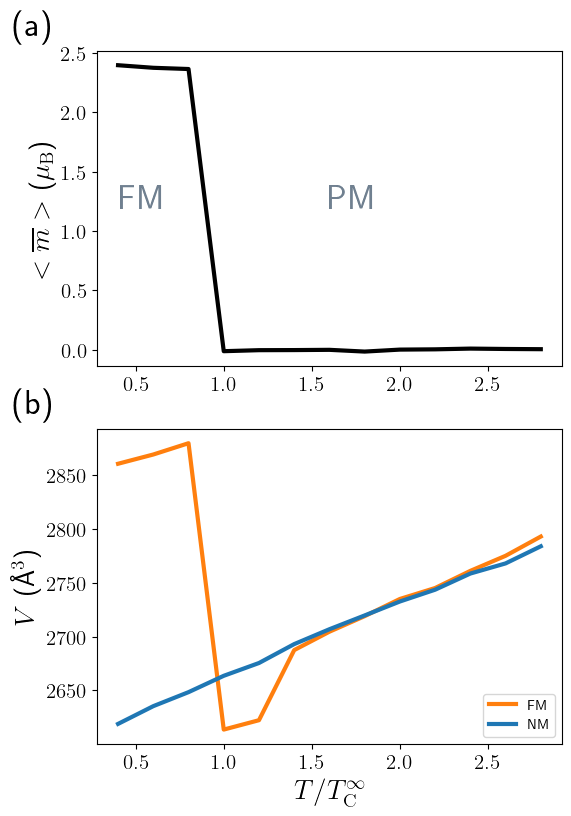}
  \caption{(a) Total magnetic moment average of a bulk Fe as a function of temperature. (b) Average linear volume expansion coefficient for Fe as a function of temperature (FM and NM states).}
    \label{fig:Figure_S2}
\end{figure}
To go further in the validation and transferability of the interatomic potential for studying thermodynamic properties, it is fundamental to assess its reliability at finite temperature. In this context, the Curie temperature is investigated and results are presented in Figure~\ref{fig:Figure_S2}a. Our TB model predicts a T$_{\mathrm{C}}^{\infty}$ around 500 K which is much lower than the experimental value of 1043 K. Improving the accuracy of the calculated Curie temperature can be done by tuning the Stoner parameter as explained in Reference~\cite{Front2022}. However, 
it turns out that properties at 0K are more difficult to reproduce in this case. Nevertheless, this deviation in the calculation of the Curie temperature does not prevent us from describing qualitatively structural and magnetic properties of Fe NPs as we will see in the following since all the results will be discussed in relation to a value of T$_{\mathrm{C}}^{\infty}$ which is simply a reference in our study. Again with the aim of studying the behaviour of our TB model at finite temperature, Figure~\ref{fig:Figure_S2}b displays the temperature dependencies of the thermal expansion coefficient in case of NM and FM calculations. A linear variation is then observed for the NM case, in contrast to the FM calculations where a contraction of the lattice parameter is reported in agreement with experiments~\cite{Liu2004}. Our TB model is therefore capable of capturing such a feature, unlike many of the interatomic potentials presented in the literature.\\


%

\end{document}